\documentstyle[epsfig,twocolumn,eqsecnum,aps]{revtex}

\begin{document}
\draft \preprint{HEP/123-qed}
\title{Corporate Default Behavior:  A Simple Stochastic Model}

\author{Ting Lei$^1$ and Raymond J.~Hawkins$^2$}
\address{$^1$Wells Fargo Bank, Capital Market Financial Products
Group, 555 Montgomery Street, San Francisco, CA 94116 \\ $^2$Bear,
Stearns Securities Corporation 220 Bush Street, Suite 845, San
Francisco, CA  94104}
\date{\today}
\maketitle
\begin{abstract}
We compare observed corporate cumulative default probabilities to
those calculated using a stochastic model based on an extension of
the work of Black and Cox \cite{blackcox76} and find that
corporations default as if via diffusive dynamics. The model,
based on a contingent-claims analysis of corporate capital
structure, is easily calibrated with readily available historical
default probabilities and fits observed default data published by
Standard and Poor's. Applying this model to the Standard and
Poor's default data we find that the difference in default
behavior between credit ratings can be explained largely by a
single variable: the ``distance to default" at the time the rating
is given.  The ability to represent observed default behavior by a
single analytic expression and to differentiate
credit-rating-dependent default behavior with a single variable
recommends this model for a variety of risk management
applications including the mapping of bank default experience to
public credit ratings.
\end{abstract}
\narrowtext
\section{Introduction}
Estimation of corporate default probability is of central
importance in credit risk management and pricing.  While
bankruptcy forecasting has been the subject of active research for
decades \cite{caouetteetal99,saunders99}, the relationship between
observed corporate default data and the stochastic dynamics of
firm value remains indirectly explored.  This is due, in part, to
the historical development of the two major research programs in
this area.  One program can be characterized as originating with
the Z-score of Altman \cite{altman68} and zeta model of Altman and
colleages \cite{altmanetal77} where a credit score is developed as
a linear function of explanatory accounting variables.  While this
approach has been successful in predicting bankruptcy and is not
inconsistent with what one might expect given the focus of rating
agencies on such financial ratios \cite{sherwood76,snp00}, its
linear deterministic structure provides limited insight into the
stochastic dynamics of the corporate default behavior. The other
major research program began with the work of Merton
\cite{merton74} who applied a contingent-claims approach
\cite{blackscholes72,merton73} to the calculation of corporate
bond spreads. Although this approach is explicitly stochastic, it
fails to predict some key features of corporate bond spreads - the
magnitude of the observed spread in general and the finite value
of the observed spread in the zero-tenor limit of the yield curve
in particular - indicating that corporate bond spread dynamics are
driven only partially by corporate default dynamics \cite{bohn99}.
Thus, while extensions of Merton's model
(e.g.~\cite{kimetal93,shimkoetal93,longstaffschwartz95}) that
typically involve multiple stochastic processes can reproduce bond
spreads, they provide limited direct information regarding the
stochastic dynamics of firm value. Nevertheless, the success these
models and of commercial products based on proprietary stochastic
models of default such as KMV's CreditMonitor$^{\rm {T}{M}}$
\cite{creditmonitor} , J.P. Morgan's CreditMetrics$^{\rm {T}{M}}$
\cite{guptonetal97}, and Moody's Public Firm Risk Model
\cite{sobehartstein00} provide compelling evidence in support of
the usefulness of a stochastic approach to explaining default
dynamics.  The purpose of our paper is to show that a simple
stochastic model of corporate default dynamics implicit in the
bond indenture work of Black and Cox \cite{blackcox76} yields a
straightforward analytic expression for the cumulative default
probability that provides a remarkably good description of the
cumulative default rates published by Standard and Poor's.

\begin{figure}
\begin{center}
\epsfig{file=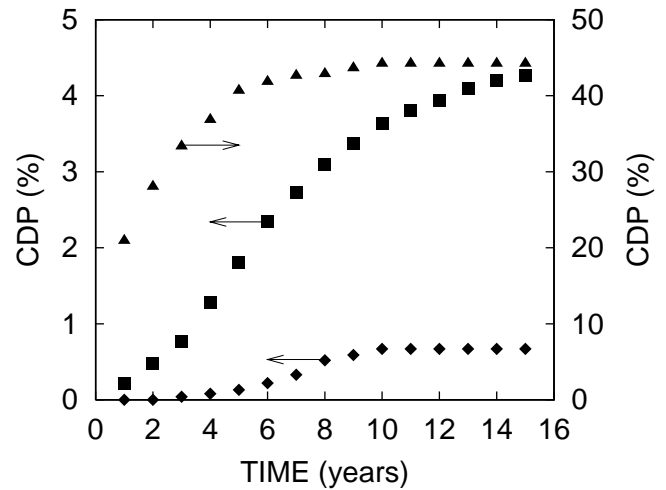}
\end{center}
\caption{Observed cumulative default probability (CDP) for AAA,
BBB, and CCC credits published by Standard and Poor's (diamonds,
squares, and triangles, respectively). The arrows point to the
appropriate y-axis for each data series.} \label{fig:abc}
\end{figure}

The empirical analysis of corporate default dynamics is
complicated by the comparatively rare nature of the corporate
default event. Fortunately, a variety of financial institutions
including rating agencies monitor and collect default data and
have compiled cumulative default probabilities.  The challenge
posed to an analytic description of corporate default is
illustrated in Figure~\ref{fig:abc} where we present the
cumulative default probability as a function of time for AAA, BBB,
and CCC rated companies published by Standard and Poor's
\cite{brandbahar00}. The AAA data denoted by the diamonds is
roughly convex for all time less than 10 years. Beyond 10 years
there are no observed defaults and the cumulative default
probability is constant.  The CCC data denoted by the triangles
are quite different with a concave function for all time.  The BBB
data show characteristics of both AAA and CCC: convex for short
times and concave for long times.  Furthermore we see that in
passing from AAA to CCC the cumulative default probabilities
change by an order of magnitude.  To describe this default
behavior we develop an analytic expression based on the structural
model of Black and Cox \cite{blackcox76} in Sec.~\ref{sec:theory}
and apply it to the observed default data published by Standard
and Poor's \cite{brandbahar00} in Sec.~\ref{sec:application}.  Our
analysis will demonstrate that a single variable in the analytic
formula provides effective discrimination between various credit
ratings and that this variable is similar to the initial
``distance to default" discussed by Crosbie
\cite{crosbie97,crosbie98}. We conclude this paper in
Sec.~\ref{sec:summary}.
\section{The Default Model}
\label{sec:theory}
The basis of a simple stochastic model of
time-dependent and credit-rating-dependent default probability
appeared years ago in the work of Black and Cox \cite{blackcox76}
where they presented a theoretical analysis of bond indenture
provisions that, among other things, examined the effect of safety
covenants on the value and behavior of corporate securities.  They
proposed a simple model of corporate capital structure where the
value of the firm $V$ would vary through time until it hit a
prescribed level.  Once $V$ reached the prescribed level indenture
agreements would specify that the firm be reorganized. Such
indenture agreements clearly impact the value of bonds and other
corporate securities as demonstrated by Black and Cox. Their work
also, however, contains the basis for a stochastic model of
bankruptcy.

The model begins with the now common assumption that a corporation
is represented as an asset with a market value, $V$, and that,
while the return of the asset is uncertain because of various
risks associated with the business, it is lognormally distributed,
namely,
\begin{equation}
\frac{dV}{V}= \mu \; dt + \sigma \; dW \; ,
\end{equation}
where $\mu$ and $\sigma$ are the constant drift and volatility of
the asset value, $t$ denotes time, and $W$ is a standard Brownian
motion.\footnote{The notion of the value of the firm as a
time-dependent stochastic variable can be traced at least as far
back as the pioneering options work of Black and Scholes
\cite{blackscholes72} and Merton \cite{merton73}, and is a basic
tenant of essentially all contingent-claims security analysis.}
Following Black and Cox \cite{blackcox76} we also assume that when
the asset value falls to a prescribed level denoted by $K$ the
company defaults. Transforming to the normalized variable $q$,
defined by
\begin{equation}
q \equiv \frac{1}{\sigma} \ln \left ( \frac{V}{K} \right ) \; ,
\end{equation}
and, using Ito's Lemma, we have that
\begin{equation}
dq = \mu^* dt + dW
\end{equation}
where $\mu^* \equiv \left( \mu / \sigma - \sigma /2 \right )$. The
default level now becomes $q = 0$.  Since $q$ is a measure of how
far the firm is from the default level, it has a natural
interpretation as the distance to default discussed by Crosbie
\cite{crosbie97,crosbie98}.  Given the initial value $q_o$, we can
calculate the expected cumulative default probability $D(t)$ from
the first-passage time probability \cite{gardiner85,ingersoll87}
\begin{equation}
D(t) = N \left ( \frac{-q_o - \mu^* t}{\sqrt{t}} \right ) + e^{-2
\mu^* q_o } N \left ( \frac{-q_o + \mu^* t}{\sqrt{t}} \right ) \;
, \label{eq:cd}
\end{equation}
where $N(x)$ is the cumulative normal distribution
function.\footnote{Those familiar with the work of Black and Cox
\cite{blackcox76} will see a strong similarity between our
Eq.~\ref{eq:cd} and their Eq.~7.  There is, indeed a direct
correspondence that can be derived by taking their reorganization
boundary to be independent of time and noting that their Eq.~7 is
for the probability that the firm has not defaulted while our
Eq.~\ref{eq:cd} is for the probability that the firm has
defaulted.} The model for ratings-based default embodied by this
expression can be interpreted as follows.  When a firm is
initially rated ($t = 0$) it will be $q_o$ standard deviations
away from default. As time passes ($t > 0$) the company's credit
state, buffeted by the vaguaries of the economic environment,
diffuses with drift $\mu^*$ and unit volatility. Should the firm's
fortunes evolve such that $q$ becomes zero, it encounters the
absorbing boundary of default. We now consider whether this model
can describe observed default behavior.
\section{The Default Model and Observed Default Behavior}
\label{sec:application}
There are two parameters in
Eq.~\ref{eq:cd}, $q_o$ and $\mu^*$, that can be varied to fit the
observed default probabilities for each rating.  As an example of
how this expression can be used to represent observed default
behavior we consider the published static pool average cumulative
default probabilities for each credit rating given by Standard \&
Poor's \cite{brandbahar00} shown in Table \ref{tab:obs}. These
values represent the probability of default as a function of time
following the {\it initial} rating of the company.  For example,
while a company that was initially rated BBB may undergo any
number of rating changes over time, once it defaults it is treated
in this analysis as being a BBB default. It can be seen that for
each rating the change in the cumulative default probability slows
(and in some cases ceases) around 10 years, reflecting the fact
that few defaults (and in some cases no defaults) have been
observed beyond 10 years.  This, as discussed below, is most
likely due to the limitation of the sample sizes available.
Consequently, we used data from the first 8 years for each rating
to fit Eq.~\ref{eq:cd}.  The parameters $q_o$ and $\mu^*$ were
obtained by minimizing the sum of the squared difference between
the observed default behavior shown in Table \ref{tab:obs} and the
calculated values obtained from Eq.~\ref{eq:cd} subject to the
constraint that the long-time cumulative default probability,
$D(\infty) = \exp \left [ -2 \mu^* q_o \right ] $, be ordered as
expected (i.e. $D_{AAA}(\infty) \leq D_{AA}(\infty) \leq \; \cdots
\; \leq D_{CCC}(\infty)$ ). This multidimensional minimization was
effected using the generalized reduced gradient (GRG2) nonlinear
optimization solver in Microsoft Excel$^{\rm TM}$.  The fitted
default probabilities using Eq.~\ref{eq:cd} and the parameters
$q_o$ and $\mu^*$ resulting from the fitting procedure just
described are shown in Table 2.

\begin{figure}
\begin{center}
\epsfig{file=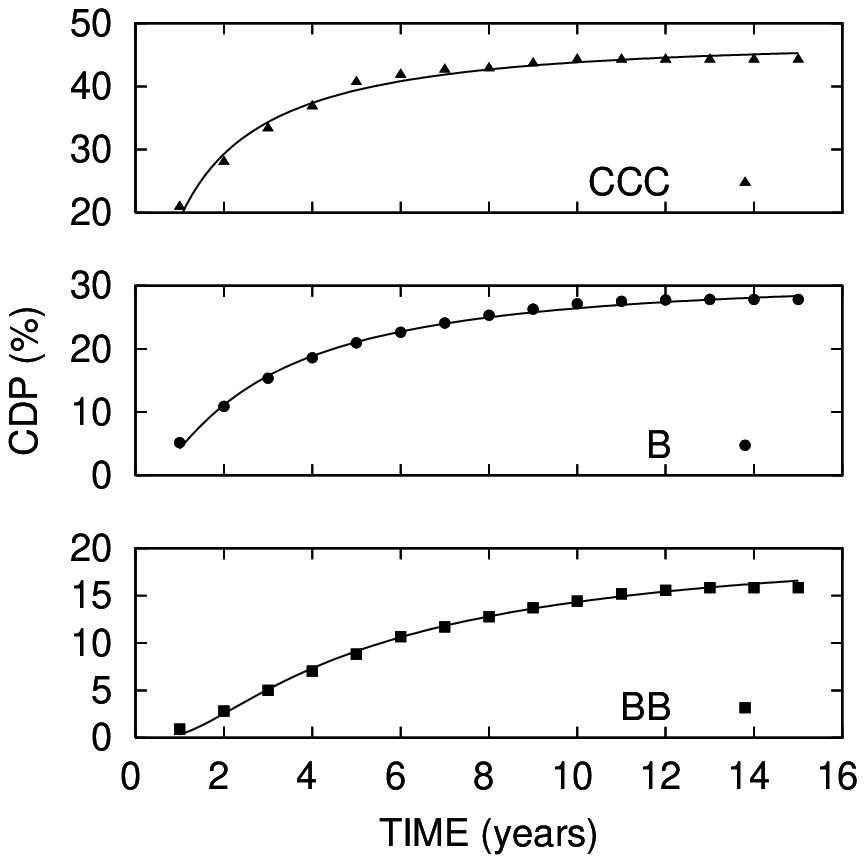}
\end{center}
\caption{Observed and calculated cumulative default probability
(CDP) for non-investment grade credits. The observed probabilities
denoted by the symbols are from the Standard and Poor's report.
The model curves are based on a fit of Eq.~\ref{eq:cd} to the
observed probabilities between 1 and 8 years.} \label{fig:nig}
\end{figure}

The result of this fit for non-investment grade credits is shown
in Table \ref{tab:calc} and Figure \ref{fig:nig}. The fit to the
CCC credit data illustrates the rather good ability of this
2-parameter model to describe the default behavior over the entire
15 year horizon based on a fit to only the first 8 years of data.
The concave nature of the data is well reproduced by
Eq.~\ref{eq:cd} as is the substantial slowing of default
probability accumulation beyond 10 years. A similarly good
description of B credits is seen in this Figure.  We see also that
BB behavior is more linear with respect to time than that of B or
CCC for horizons less than 5 years and that the model is able to
follow this as well as the more concave behavior between 5 and 13
years. Beyond 13 years there are no observed defaults and the
cumulative default probability is that same from year 13 to year
15 because of this. The model, however, continues to rise
gradually indicating that the lack of observed defaults is,
likely, due to a limited sample size and that, in time, we can
expect to see more defaults in this area. Financially, there is
nothing special about the 13$^{th}$ year after a BB rating that
would account for the observed sudden lack of defaults.

\begin{figure}
\begin{center}
\epsfig{file=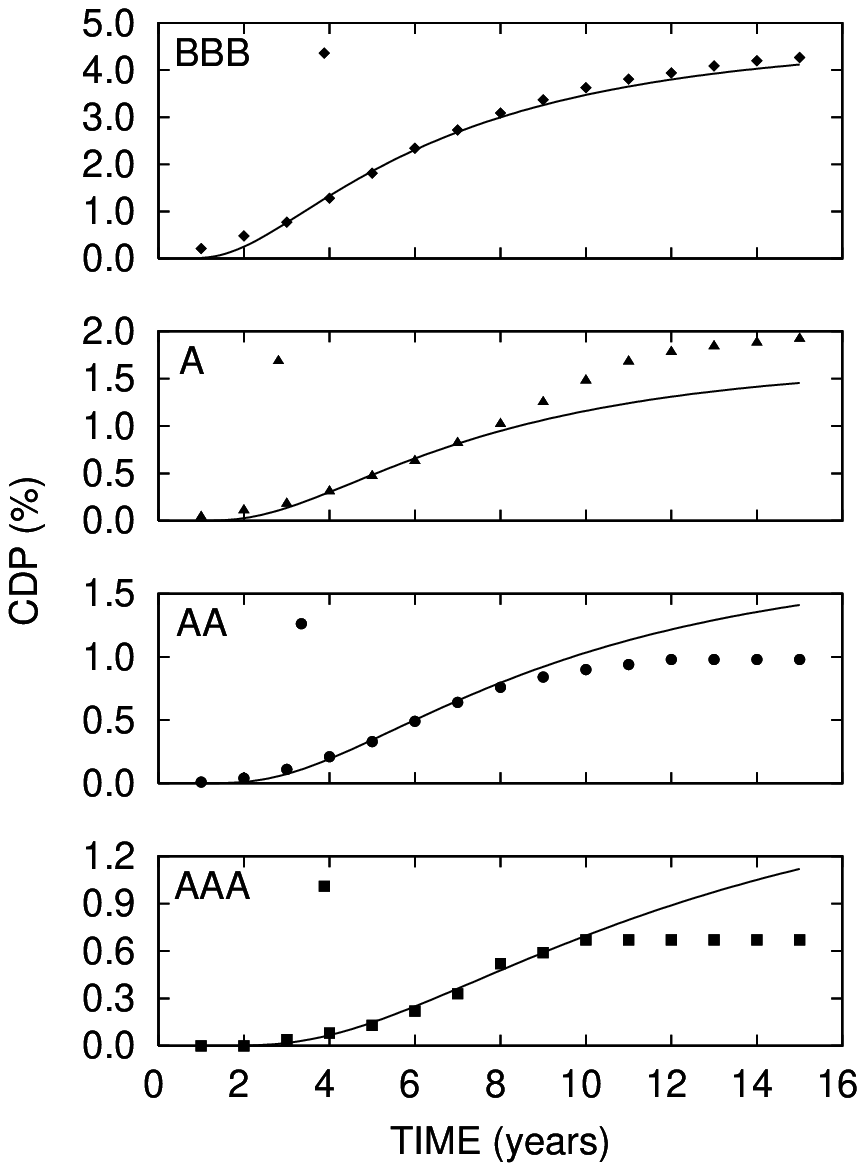}
\end{center}
\caption{Observed and calculated cumulative default probability
(CDP) for investment grade credits. The observed probabilities
denoted by the symbols are from the Standard and Poor's report.
The model curves are based on a fit of Eq.~\ref{eq:cd} to the
observed probabilities between 1 and 8 years.} \label{fig:ig}
\end{figure}

The result of our fit for investment grade credits is shown in
Table \ref{tab:calc} and Figure \ref{fig:ig}. The BBB credits
shown in this Figure illustrate an interesting qualitative change
in default behavior: the short horizon data are {\it convex} in
time. The intermediate horizon data are linear in time and the
longer horizon data are concave. We see that our simple
2-parameter model provides a very reasonable description of the
data. The cumulative default behavior A credits is a challenge to
the model. The model clearly tracks the observed data well between
1 and 8 from years and 11 to 15 years, but an accumulation of
defaults in the 8 to 11 year time frame results in a substantial
offset between the observed and the calculated data in the 11 to
15 year region.  One solution is to fit over the 1$^{st}$ 10
years.  While this is perfectly reasonable and most people using
such a model to fit their data would likely do so, we felt it
useful to see how far we could get using a uniform time range for
fitting purposes. For AA and AAA credit default behavior we see
good fits to observations over the 1$^{st}$ 8 years and reasonable
extrapolations with significant deviations coinciding with the
point at which the cumulative default data stop changing due to
lack of observed defaults.

\begin{figure}
\begin{center}
\epsfig{file=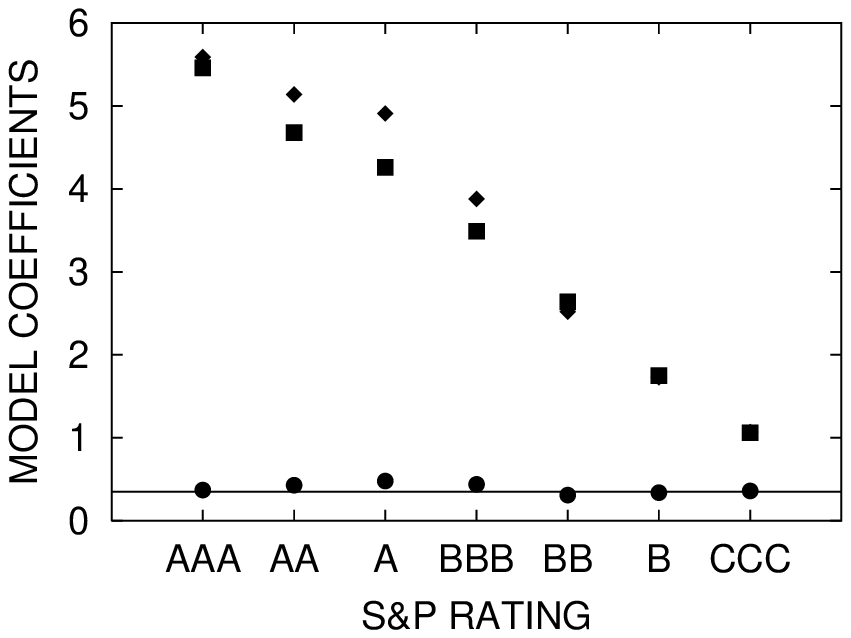}
\end{center}
\caption{The calculated initial distance to default $q_o$
(daimonds and squares) and drift $\mu^*$ (circles and horizontal
line) as a function of credit rating.} \label{fig:coeff}
\end{figure}

The coefficients of Eq~\ref{eq:cd}, $q_o$ and $\mu*$, that
resulted from the fitting procedure that generated Figures
\ref{fig:nig} and \ref{fig:ig} are shown as a function of credit
rating in Figure \ref{fig:coeff}.  The squares correspond to the
fitted values of $q_o$ and the circles correspond to the fitted
values of $\mu^*$.  This graph demonstrates the intuitively
expected result that the better the credit rating the larger the
initial distance to default. Recalling from our earlier discussion
that the distance to default is measured in standard deviations,
we see that the average AAA company is initially about 5.5
standard deviations from default, the average BBB company is
initially about 4 standard deviations from default, and the
average CCC company is initially about 1 standard deviation from
default.

The normalized drift, $\mu^*$, resulting from the fitting
procedure that generated Figures \ref{fig:nig} and \ref{fig:ig} is
remarkable in that, while each credit was fit individually, the
normalized drifts are quite similar. This similarity prompted us
to explore the results that would follow if $\mu^*$ was assumed
{\it a priori} to be the same for all credits. The results of a
global fit with this restriction are shown as diamonds in Figure
\ref{fig:coeff} where we see the initial distance-to-default $q_o$
as a function of credit. This analysis also found that $\mu^* =
0.35$: shown as a horizontal line in Figure \ref{fig:coeff}.
Comparing the $q_o$ for constant and variable drift we see that
setting $\mu^*$ constant across credits has essentially no impact
on $q_o$ for the lower credits and has a minor impact on the
higher credits.  That the default dynamics of all credits can be
represented by a single value of $\mu^*$ implies that differences
in cumulative default behavior among various ratings are driven
almost exclusively by the initial distance to default $q_o$.

Modeling the default process as a first-passage time yields a
simple expression for the mean time to default:  $q_o / \mu^*$.
Comparing the results of this expression with those reported by
Standard and Poor's in Table \ref{tab:mtd}.  The deviation between
the calculated and observed results reflects the lack of observed
default at longer tenors.  However, for the same reason that we
would expect the longer-tenor cumulative default probability for
the AA and AAA credits to increase over time, so too do we expect
the mean time to default to increase over time for
investment-grade credits.
\section{Summary}
\label{sec:summary} Comparing observed corporate cumulative
default probabilities to those calculated using a stochastic model
based on an extension of the work of Black and Cox
\cite{blackcox76}, we find that corporations default as if via
diffusive dynamics.  The model, based on a contingent-claims
analysis of corporate capital structure, yields a single analytic
expression for corporate default behavior that is calibrated
easily with historical default probabilities. We used this model
to analyze the observed default data published by Standard and
Poor's \cite{brandbahar00} and found that a single variable in the
analytic formula provides effective discrimination between various
credit ratings.  This variable is quite similar to the ``distance
to default" described by Crosbie \cite{crosbie97,crosbie98} and
provides an attractive interpretation of the default process in
terms of the bond indenture provision analysis of Black and Cox
\cite{blackcox76}. Despite its simple underpinnings, the model is
remarkably successful in describing the cumulative default rates
published by Standard and Poor's \cite{brandbahar00}.  This
implies that the capital structure of corporations, despite their
differences, map onto the simple ``effective" capital structure
given in the Merton model. The ability to represent observed
default behavior by a single analytic expression and to
differentiate credit-rating-dependent default behavior with a
single variable recommends this model for a variety of risk
management applications including the mapping of bank default
experience to public credit ratings.

\acknowledgments

We thank Arden Hall and Jim Westfall for enlightening discussions
and Vincent Chu and Terry Benzschawel for their helpful comments.
This article was written before Ting Lei joined Wells Fargo and
Raymond J. Hawkins joined Bear, Stearns Securities Corporation.
Neither Wells Fargo or Bear, Stearns Securities Corporation are
responsible for any statements or conclusions herein; and no
opinions, represented herein in any way represent the position of
Wells Fargo or of Bear, Stearns Securities Corporation.


\begin{table}
\caption{Observed cumulative default probabilities (in percent) as
reported by Standard \& Poor's.} \label{tab:obs}
\begin{tabular}{cccccccc}
  Time (years) & AAA & AA & A & BBB & BB & B & CCC \\
  1 & 0.00 & 0.01 & 0.04 & 0.21 & 0.91 & 5.16 & 20.93 \\
  2 & 0.00 & 0.04 & 0.11 & 0.48 & 2.82 & 10.90 & 28.04 \\
  3 & 0.04 & 0.11 & 0.18 & 0.77 & 5.00 & 15.36 & 33.35 \\
  4 & 0.08 & 0.21 & 0.31 & 1.28 & 7.04 & 18.60 & 36.83 \\
  5 & 0.13 & 0.33 & 0.47 & 1.81 & 8.82 & 20.95 & 40.67 \\
  6 & 0.22 & 0.49 & 0.63 & 2.34 & 10.68 & 22.65 & 41.83 \\
  7 & 0.33 & 0.64 & 0.82 & 2.73 & 11.71 & 24.08 & 42.64 \\
  8 & 0.52 & 0.76 & 1.02 & 3.09 & 12.78 & 25.32 & 42.86 \\
  9 & 0.59 & 0.84 & 1.25 & 3.37 & 13.71 & 26.29 & 43.63 \\
  10 & 0.67 & 0.90 & 1.48 & 3.63 & 14.42 & 27.13 & 44.23 \\
  11 & 0.67 & 0.94 & 1.68 & 3.81 & 15.19 & 27.54 & 44.23 \\
  12 & 0.67 & 0.98 & 1.78 & 3.94 & 15.55 & 27.76 & 44.23 \\
  13 & 0.67 & 0.98 & 1.84 & 4.09 & 15.84 & 27.83 & 44.23 \\
  14 & 0.67 & 0.98 & 1.88 & 4.20 & 15.84 & 27.83 & 44.23 \\
  15 & 0.67 & 0.98 & 1.92 & 4.27 & 15.84 & 27.83 & 44.23 \\
\end{tabular}
\end{table}

\begin{table}
\caption{Calculated cumulative default rates (in percent) obtained
by fitting Eq.~\ref{eq:cd} to the data in Table \ref{tab:obs}.}
\label{tab:calc}
\begin{tabular}{cccccccc}
  Time (years) & AAA & AA & A & BBB & BB & B & CCC \\
  1 & 0.00 & 0.00 & 0.00 & 0.01 & 0.35 & 4.24 & 19.19 \\
  2 & 0.00 & 0.01 & 0.03 & 0.25 & 2.54 & 11.11 & 29.49 \\
  3 & 0.02 & 0.07 & 0.14 & 0.77 & 5.08 & 15.72 & 34.60 \\
  4 & 0.07 & 0.19 & 0.31 & 1.35 & 7.27 & 18.84 & 37.66 \\
  5 & 0.15 & 0.34 & 0.50 & 1.88 & 9.06 & 21.06 & 39.69 \\
  6 & 0.25 & 0.49 & 0.67 & 2.34 & 10.51 & 22.71 & 41.13 \\
  7 & 0.36 & 0.64 & 0.83 & 2.73 & 11.70 & 23.97 & 42.20 \\
  8 & 0.47 & 0.78 & 0.97 & 3.05 & 12.68 & 24.97 & 43.02 \\
  9 & 0.58 & 0.90 & 1.09 & 3.31 & 13.49 & 25.76 & 43.66 \\
  10 & 0.69 & 1.01 & 1.19 & 3.53 & 14.17 & 26.41 & 44.17 \\
  11 & 0.79 & 1.10 & 1.27 & 3.72 & 14.76 & 26.94 & 44.59 \\
  12 & 0.88 & 1.19 & 1.35 & 3.87 & 15.25 & 27.38 & 44.93 \\
  13 & 0.96 & 1.26 & 1.41 & 4.00 & 15.68 & 27.76 & 45.22 \\
  14 & 1.04 & 1.32 & 1.46 & 4.11 & 16.05 & 28.07 & 45.46 \\
  15 & 1.11 & 1.37 & 1.50 & 4.20 & 16.37 & 28.35 & 45.66 \\
\end{tabular}
\end{table}

\begin{table}
\caption{Calculated and observed mean time to default(in years). }
\label{tab:mtd}
\begin{tabular}{cccc}
  Rating & Variable $\mu*$ & Constant $\mu*$
  &Observed \\
  AAA   & 14.7  & 16.1  & 8.0 \\
  AA    & 10.8  & 14.8  & 8.3 \\
  A     & 9.0   & 14.1  & 8.2 \\
  BBB   & 8.0   & 11.2  & 6.6 \\
  BB    & 8.4   & 7.2   & 4.7 \\
  B     & 5.1   & 5.0   & 3.4 \\
  CCC   & 3.0   & 3.1   & 3.2   \\
\end{tabular}
\end{table}

\end{document}